# Transverse localization of light in 1D disordered waveguide lattices with backbone photonic bandgap

**Somnath Ghosh[1]\*, B. P. Pal[2], R. K. Varshney[2] and H. Ahmad[3]**

[1] Institute of Radio Physics and Electronics, University of Calcutta, Kolkata 700009, India
[2] Physics Department, Indian Institute of Technology Delhi, Hauz Khas, New Delhi 110016, India
[3] Photonics Research Centre, University of Malaya, 50603 Kuala Lumpur, Malaysia

Email: somiit@rediffmail.com*

**Abstract**
The role of a prominent photonic bandgap (PBG) on the phenomenon of transverse localization of light in a semi-infinite lossless waveguide lattice consisting of evanescently coupled disordered one-dimensional optical waveguides has been investigated numerically. The interplay between the underlying photonic bandgap due to inherent periodicity of the optical system and various levels of deliberately induced transverse disorder in its refractive index periodicity has been studied. We show that the PBG indeed plays an important role and its simultaneous presence could catalyze realization of localized light even when strength of disorder is not sufficiently strong to independently cause localization of light. An important outcome of this study revealed that PBG could be gainfully exploited to tailor the spectral window for localization of light in potential applications like lasing in a disordered optical lattice.

**Keywords** : Localization of light, disordered waveguide lattices, photonic bandgap

## 1. Introduction

De Raedt et al [1] re-visited the idea of light localization [2] and introduced the concept of transverse light localization in a semi-infinite disordered geometry in which light confinement occurs only in a plane perpendicular to the direction of light propagation. This interesting phenomenon of light localization in 1D/2D disordered dielectric structures analogous to Anderson localization, which proved to be experimentally realizable, has emerged as a research field of intense contemporary interest [3, 4, 5]. It is now established that in temporarily or permanently realized lattices with deliberately introduced disorder, light confinement occurs due to the sole effect of disorder [5-9]. The backbone periodic structures in these disordered geometries are likely to influence (unlike a completely random structure) the phenomenon of localization of light as a combined effect of long range order and disorder. Ongoing intensive research on photonic bandgap structures, deal with confinement of light within a localized defect region in an otherwise periodic structure in which the periodicity accounts for the photonic bandgap. Further, with the advent of discrete photonic systems, this new avenue of light confinement has emerged as a contemporary field of research in the context of disordered optical structures [7, 10].

However, with the state-of-the art fabrication process for developing such photonic structures, having characteristic features, which are of micrometer/ sub-micrometer scale, it is not an easy task to realize a perfectly periodic structure in practice. In view of this, in our opinion transverse localization (TL) of light has added a new dimension to light guidance in PBG structures. Hence, due to this unwanted deviation, the migration of new accessible states from the band edge towards the centre of the inherent bandgap of a targeted periodic structure has become important. This effect eventually smears out the desired effect of decreased density of states (DOS) because of filling the PBG by band-tail localized states [11]. Hence, investigation on the stability of the photonic bandgap in the presence of disorder is an important aspect even from the point of view of estimating the quality of an intended PBG in an optical structure [12,13]. Thus, simultaneous presence of PBG and disorder, essentially synthesizes two distinctly different phenomena namely, PBG guidance and localization of light has become an important problem, which should be of great contemporary interest from the application as well as pure theoretical point of view.

In the above mentioned studies [4, 5, 6] on the localization effect, though there existed an underlying long range periodicity, the significance of a prominent bandgap in the context of TL of light has not been systematically investigated in the published literature. In an optical geometry having a bandgap, the phenomenon of light localization due to the inherent Bragg scattering by the ordered structure is fundamentally different from the effect of TL in the presence of a disorder. This could be the reason for absence of such a study. This motivated us to investigate the phenomenon of TL of light in the presence of a prominent bandgap.

## 2. Modeling of disordered lattices

In this paper, we study the influence of the existence of a prominent photonic bandgap on the controlled disorder-assisted TL of light in a 1D waveguide lattice. The sample lattice geometry is chosen to be such that at the operating wavelength there exists a prominent PBG and it is assumed that disorder is introduced into it in the form of a perturbed refractive index in transverse direction. Our results show that even only a small disorder (of much lower strength than the threshold disorder [9,15] required for the case of solely disorder-assisted TL in absence of PBG) is sufficient to achieve localization in the simultaneous presence of a bandgap. Also, one may perhaps exploit this by appropriately choosing the lattice parameters as an additional tool to tune the PBG along with the embedded disorder to attain spectral selectivity of the localization phenomenon in specific applications like disordered/ random [14] lasing.

We consider an evanescently coupled waveguide lattice consisting of a large number ($N$) of unit cells, and in which all the waveguides spaced equally apart are buried inside a medium of constant refractive index $n_0$ [4,6,9]. The overall structure is homogeneous in the longitudinal ($z$) direction along which the optical beam is assumed to propagate. The change in refractive index $\Delta n(x)$ (over the uniform background of $n_0$) due to disorder in this 1D waveguide lattice is assumed to be of the form

$$\Delta n(x) = \Delta n_p (H(x) + C\delta(x)) \quad (1)$$

here $C$ is a dimensionless constant, whose value governs the level/strength of disorder; the periodic function $H(x)$ takes the value 1 inside the higher-index regions and is zero elsewhere; $\Delta n(x)$ consists of a deterministic periodic part $\Delta n_p$ of spatial period $\Lambda$ and a spatially periodic random component $\delta$ (uniformly distributed over a specified range varying from 0 to 1). This particular choice of randomly perturbed refractive index in the high index as well as low index layers enables us to model the diagonal and off-diagonal disorders to study the localization of light [4, 6, 9]. Wave propagation through the lattice is governed by the standard scalar Helmholtz equation, which under paraxial approximation can be written as

$$i\frac{\partial A}{\partial z} + \frac{1}{2k}(\frac{\partial^2 A}{\partial x^2}) + \frac{k}{n_0}\Delta n(x) A = 0 \quad (2)$$

where A(x,z) is amplitude of an input CW optical beam having its electric-field as

$$E(x,z,t) = \text{Re}[A(x,z)e^{i(kz-wt)}]; \; k = n_0\omega/c$$

To study the effect of a PBG on the phenomena of transverse localization of light, one could write, for the lattice, the following equation:

$$\left.\begin{array}{l}\Delta n(x) = \Delta n_p (H(x) + C\partial(x)); C \neq 0 \\ d_1 + d_2 = \Lambda\end{array}\right\} \quad (3)$$

where the Eq. (3) represents the index [4,9,15,16] disorder only, unlike spatial disorder [17,18]. We have also deliberately excluded the simultaneous presence of spatial as well as index disorder in the context of this investigation. We solve Eq. (2) numerically through the scalar beam propagation method, which we have implemented in Matlab. In our optimized 10 $mm$ long 1D lattice geometry of 150 identical evanescently coupled waveguides, we consider the high index ($n_2$) regions of width ($d_2$) 3 μm, which are separated by low index regions ($n_1$) equal distances ($d_1$) of 6 μm. The value of $\Delta n_p$ was chosen to be 0.016 over the background material of refractive index $n_0$ = 1.454. We have chosen this particular distribution of the transverse refractive index of the perfectly ordered lattice such that it spawns a prominent bandgap.

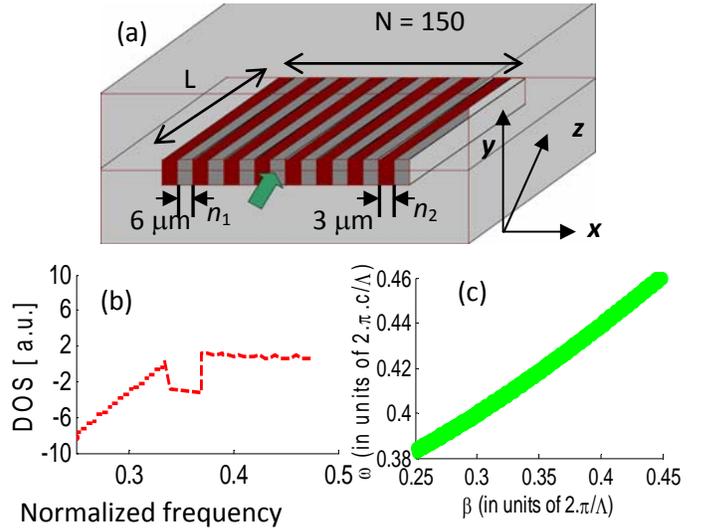

**Figure 1.** (Color online) a) Schematics of the chosen 1D coupled waveguide ordered lattice. b) Calculated density of states (DOS) and c) lowest order stopband of the lattice.

In Fig. 1(a) we have depicted the schematic of the designed lattice geometry without any disorder introduced, whereas Fig. 1(b) shows the numerically estimated density of states plot in the presence of the PBG due to a perfectly periodic structure. The extent of the corresponding fundamental bandgap in the operating frequency scale (ω) of the lattice has also been shown in Fig. 1(c) where β is the propagation constant of eigen mode. Once the band-structure is obtained, before introducing the different levels of refractive index perturbations in the lattice we have identified the wavelength window of the stop band and accordingly three important regimes of operation have been chosen: one well inside the bandgap, one just outside its stop-band (i.e. near the band edge) and one well outside the bandgap, respectively. With these suitable choices of the operating wavelengths, it should be possible to investigate the interplay between transverse disorder and photonic bandgap from the point of view of localization.

## 3. Results and discussions

In order to investigate quantitatively the effect of having a background photonic bandgap on its localization in a disordered medium, we have studied the beam dynamics (with the input Gaussian beam of FWHM 8 μm) for different lengths of the lattice at different operating wavelength regimes. A measure of the localization is assumed to be quantifiable through decrease in the average effective width ($\omega_{eff}$) (as defined in) [4,15]

$$P \equiv \left[\int I(x,L)^2 dx\right] / \left[\int I(x,L) dx\right]^2$$

$$\omega_{eff} = \langle P \rangle^{-1} \quad (4)$$

of the propagating beam after including the statistical nature of the localization phenomenon in a finite system; where <...> represents a statistical average over several realizations of the same level of disorder. In Fig. 2(a), we have shown the spectral dependence of $\omega_{eff}$. This is a key parameter which is well defined (*unlike the localization length which is defined only in the localized regime*) in both ballistic as well as localized regime of beam dynamics to identify the nature of particular eigenmodes of the lattice and which is directly proportional to the localization length of the state in the localized regime [4, 9, 15, 16] around the center of the fundamental bandgap for a chosen lattice length of 10 mm when the values of $C$ are set at 0 (absence of disorder), and 0.40, respectively. This particular plot clearly shows the signature of the presence of a PBG (when $C = 0$). As we increase $C$ to 0.40, the consequent disorder destroys the bandgap effect and the wavelength selectivity of $\omega_{eff}$ inside the lattice almost disappears as could be seen from Fig. 2(a). essentially, the bandgap loses its prominence when more number of accessible states appear inside the bandgap and the corresponding $\omega_{eff}$ variation near the edge becomes relatively flat. The variations in spectral derivatives of the $\omega_{eff}$ corresponding to four different strengths of disorder are depicted in Fig. 2(b). The trend in these variations clearly manifests the existence of a boundary (near λ = 1020 nm) at the band center between two distinct categories of modes inside a PBG. As we increase the magnitude of $C$, in the structure the two prominent peaks around the band center diminishes and the overall behavior becomes almost flat. Physically, the ordered lattice loses its standing wave-like feature inside the bandgap as we increase the $C$ parameter to a value > 0.25 and it forms a band consisted of localized states ($C = 0.40$). Therefore, there exist a critical level of disorder for a given lattice above which signature of the underlying photonic bandgap is completely destroyed. The disorder takes over to control the nature of the states (standing-wave like state due to interference effect inside the bandgap to a localized state) to form band.

To appreciate the after-effect of the deliberate disorder in a lattice in the presence of a PBG, we have chosen certain wavelengths from Fig. 2(b) in three different regimes in and around the bandgap. We first choose a wavelength near the center of the PBG (λ = 1020 nm), then one well outside the PBG (λ = 1400 nm), and finally one near the band edge (λ = 1060 nm) respectively. To quantify the effect of disorder in the waveguide lattice, we introduce $C$ from 0 to 0.50 upwards in steps of 0.10 and investigate the beam dynamics inside the lattice. In the simultaneous presence of disorder and PBG, the effect of bandgap is gradually reduced as we increase the level of disorder and at the same time the sole effect of disorder in terms of localization effect surfaces. The dynamics of this subtle interplay between PBG effect and localization effect is different at different operating wavelength regimes with respect to the bandgap. Results in terms of spectral dependence of $\omega_{eff}$ which contains crucial information regarding the spectral properties of the natural states of the lattices are shown in Fig. 3. It can be seen from Fig. 3 that when we operate well inside the bandgap (i.e. $\lambda = 1020$ nm), the propagating beam get localized due to the bandgap effect (at $C = 0$) and evolves to a state with relatively smaller value for the $\omega_{eff}$. Inside the PBG, a Bloch state with its exponential envelope covers relatively less number of lattice units. However, as we introduce disorder, following the trend as shown in Fig. 2 the influence of bandgap becomes less which results in a relatively poor confinement of the output beam (upto $C = 0.10$). Thus for $C$ upto 0.10, the beam width increases (unlike the signature of localization) [4, 9]. However as $C$ is increased beyond this value, disorder-assisted localization effect dominates over the interplay between bandgap and disorder and the light beam get localized for a $C \geq 0.25$. Beyond that point the variation of $\omega_{eff}$ follow the universal behavior of TL i.e. $\omega_{eff}$ decreases with increase in $C$. It could be concluded from these results that disorder overtakes bandgap effect at $C$ around 0.10. Hence it reveals that the transition from a ballistic mode of propagation to a localized mode in a disordered lattice without having a PBG [9] is smoother than the transition from a bandgap guided state to a localized state. In other words, in the presence of a prominent bandgap, localization of light would occur at a relatively low level of disorder contrary to its counterpart case in which PBG is absent. Thus essentially presence of a PBG would act as a catalyst to realize localization at a relatively weak disorder. With this unique feature these lattices could be one of the best candidate/ platform to achieve localized light. Even from the point of view of spectral control of the TL effect, these lattices are more suitable. An interesting behavior is observed when the operating condition is chosen near the band edge (i.e. $\lambda = 1060$ nm). In a perfectly ordered lattice the edge states are always localized. They carry signatures of both bandgap guidance (similar to exponentially localized states within the bandgap) and ballistic mode of propagation (similar to extended states outside the bandgap in the absence of disorder) as they correspond to intermediate states. Hence this particular class of localized modes exhibits the characteristic decaying tail at a rate slower than exponential. When disorder is introduced, the spectral window of bandgap is slightly enhanced but with reduced efficiency (as shown in Fig. 2), subsequently the combined effect of disorder and PBG [12, 19] and sole effect of transverse disorder together play the role to control the nature of these states. When the level of disorder is not sufficiently strong (i.e.

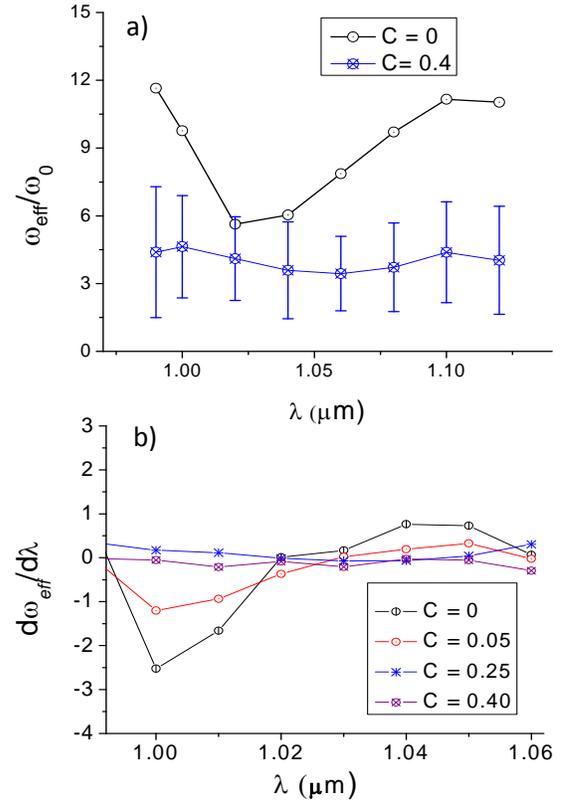

**Figure 2.** Variation in a) ensemble averaged effective width of the output beam ($\omega_{eff}$) inside the PBG for two different C values, C = 0, and 0.4 respectively; b) spectral derivative of ($\omega_{eff}$) with the operating wavelength for an input Gaussian beam (FWHM 8 μm) for three different strengths of disorder (C = 0.05, 0.25 and 0.4), other than the case of perfectly ordered lattice after propagation through 10 mm sample lengths. The range of wavelength considered (x-axis) is dictated by the backbone PBG of the lattice. The error bars indicate the statistical standard deviations (ssd) of beam widths for 100 realizations.

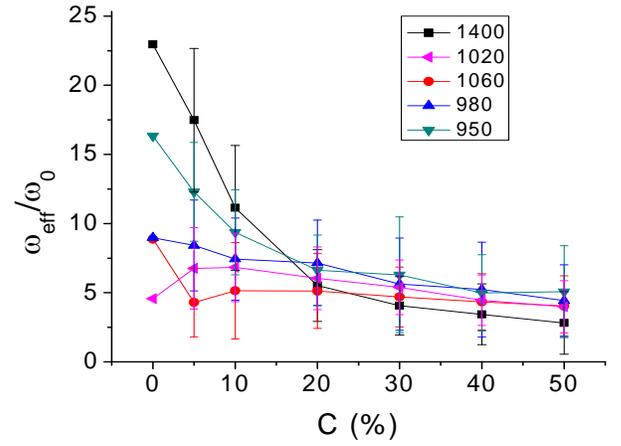

**Figure 3.** (Color online) Variation in the ensemble averaged effective width (along with the ssd) of the output beam for an input Gaussian beam (FWHM 8 μm) for various levels of disorder after propagation through 10 mm. The figure shows the interplay between the refractive index disorder and PBG.

relatively small $C$); the influence of underlying bandgap is reduced though still exists while TL effect is visible as the ballistic feature of propagation becomes less prominent. In this regime as we increase $C$ (~ 0.05), the beam width gets reduced rapidly. However, when a stronger level of disorder ($C \geq 0.2$) is chosen, the bandgap effect is destroyed and the modes essentially exhibit signature of TL. In this regime as we increase C, the beam width also decreases. In between for moderate values of $C$, a state of transition could be seen, which is marked by the peak (near $C = 0.10$) in the variation of $\omega_{eff}$ with $\lambda$.

Thus the above results of our study for a 1D waveguide array forming a disordered lattice clearly show that there exist two quantitatively different regimes of localization if PBG exists as a backbone. States outside the PBG and near the band edge are categorized as regime-I, which follow the normal behavior of TL, whereas the regime-II consists of the states well inside the PBG. This particular regime shows the anomalous behavior unlike TL below a certain critical value of disorder. However, above that critical level of disorder both the regimes follow normal behavior according to the well known signature of TL. It can be seen that $\omega_{eff}$ decreases with increase of disorder at high enough $C$ in both the regimes. As the overall behavior of Fig. 2 is directly related to the bandgap feature of the parent periodic structure, the transition of $\omega_{eff}$ variation with $C$ (as shown in Fig. 3) from one shape to another occurs within a narrow spectral band.

For a deeper appreciation of this interplay between PBG and disorder on the degree of localization, we have plotted in Fig. 4(a) the ensemble averaged intensity profiles (averaged over 100 output intensity profiles) at the output end of a 10 mm long lattice; the operating wavelength was chosen to be near the center of the PBG and C values were set at 0 and 0.05, respectively. The output profile from the periodic lattice carries the signature of localization due to PBG. In this case light is exponentially localized due to the presence of imaginary part of the Bloch wave-vector and it covers relatively lesser number of lattice units. With C set at 0.05, the transition from a bandgap influenced state to a localized state under the combined influence of disorder and PBG is evident. Hence the $\omega_{eff}$ slightly increases and intensity profile eventually acquires two linearly decaying tails when plotted on a semi-log scale in Fig. 4(b); which is indeed the hallmark of transverse localization. Also this state covers a relatively large number of lattice units compared to a PBG-assisted state.

To appreciate the interesting behavior (as discussed in Fig. 2) of competition between disorder and PBG near the band edge (i.e. at λ = 1060 nm) we have studied the ensemble-averaged (over 100 realizations) output intensity profiles from the 10 mm long waveguide lattice when the level of disorder is set at C = 0, 0.05 and 0.10, respectively. The localized edge states in the periodic structure (C = 0) are states intermediate between the ballistic and bandgap guided (Bloch) states as they carry the characteristic feature of both the states (ballistic side lobes with slowly decaying tails). These localized edge states are very sensitive to disorder, which affects PBG effect and at the same time favors localization. When C is set to be 0.05, we have achieved a state in which the ballistic nature of the tails is now less prominent and simultaneously the exponentially decaying nature (due to combined effect of PBG with slightly extended spectral window and TL) dominate. But as we further increase C beyond 0.10, the bandgap effect dies off and dominance of transverse disorder becomes evident.

Until now, we have elaborated the interplay between a PBG and the presence of a deliberate and controlled transverse disorder and established that the phenomena of TL exhibits certain interesting features in the presence of a bandgap, which are not discernable in the absence of PBG. In the absence of PBG, propagation dynamics in the lattice is solely dictated by long-range disorder in lower dimensions. However, interesting additional features are revealed when we compare the behavior of the states in different localized regimes (well inside the bandgap, near the band edge and far away from the bandgap). In Fig. 5(a) we have re-plotted the beam dynamics with various levels of disorder for two particular cases near the band edge (λ = 1060 nm) and value at λ = 1060 nm. To visualize the contribution of the underlying bandgap of the chosen lattice to the localized states as a result of combined effect of PBG and disorder, we introduce an isolated defect (no long range correlation in spatial dimension of the lattice index profile) waveguide in the lattice near the central far away from the gap (λ = 1400 nm). From this figure, it is evident that the localization effect at the operating wavelength of 1400 nm takes over the localization behavior at 1060 nm for C beyond 0.25. With any further increase in C, $\omega_{eff}$ approaches smaller values at the operating wavelength of 1400 nm as compared to its region (77[th] high index region). As the transverse disorder is in the form of perturbed refractive index, the defect waveguide was chosen to be of a spatial width of 3.7 μm, different from 3 μm chosen for all other waveguides (high index regions). This particular choice of the defect introduces a defect state inside the bandgap of the lattice. Accordingly, we assume launch of two different wavelengths (λ = 1060 and 1400 nm respectively) into the defect waveguide. In the presence of this defect inside the lattice, we plot the ensemble averaged output profiles for $C$ as 0.25 (as shown in Fig. 5(b)) and 0.40 (as shown in Fig. 5(c)), respectively in the respective localized regimes (as shown in Fig. 5(a)). In both the cases, while operating at 1060 nm, it excites a localized defect state, which is more confined compared to its counterpart defect state. This additional confinement factor of the localized defect state is attributed to the bandgap effect (only present at λ = 1060 nm) in addition to the sole effect of disorder (which is present both at λ = 1060 nm and 1400 nm, respectively) to realize these localized states. A direct comparison between the states plotted in Fig. 4(a) and those in Figs. 5(b) and (c), confirms that the localized defect states at 1060 nm is more strongly localized than the pure disorder-induced transverse state or a pure bandgap state. It may be noted that a very similar behavior has been observed while the operating wavelength was chosen very close to the center of the PBG (i.e. at λ = 1020 nm).

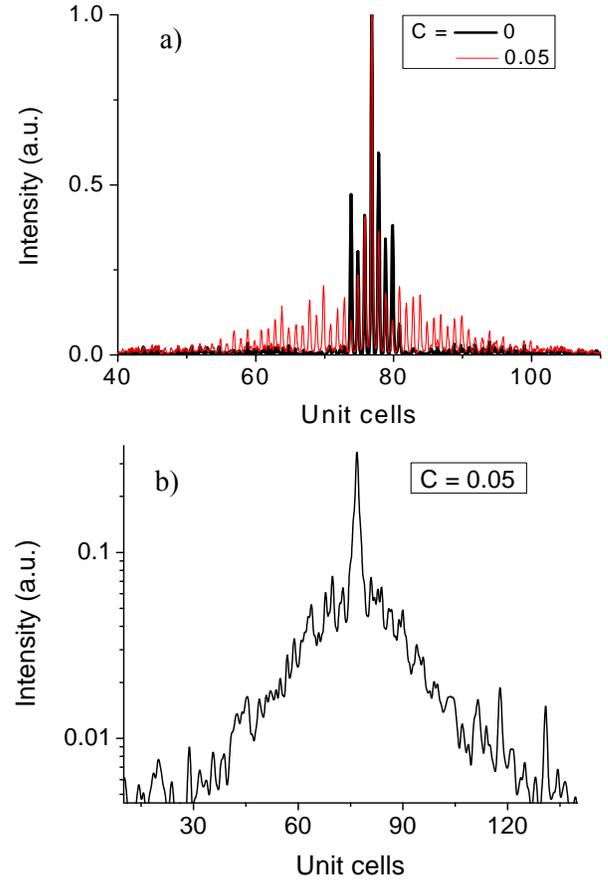

**Figure 4.** (Color online) a) Ensemble averaged output intensity profiles for the case of absence of disorder (in black) and a deliberate disorder of 5% (in red) from a 10 mm long lattice, respectively, when a Gaussian beam (FWHM 8 μm) was injected at the input of wavelength 1020 nm. b) Output intensity profile on a semi-log scale shows a nearly linear variation in the tails, thereby manifesting the signature of transverse localization in a partially disorder lattice in the presence of a PBG.

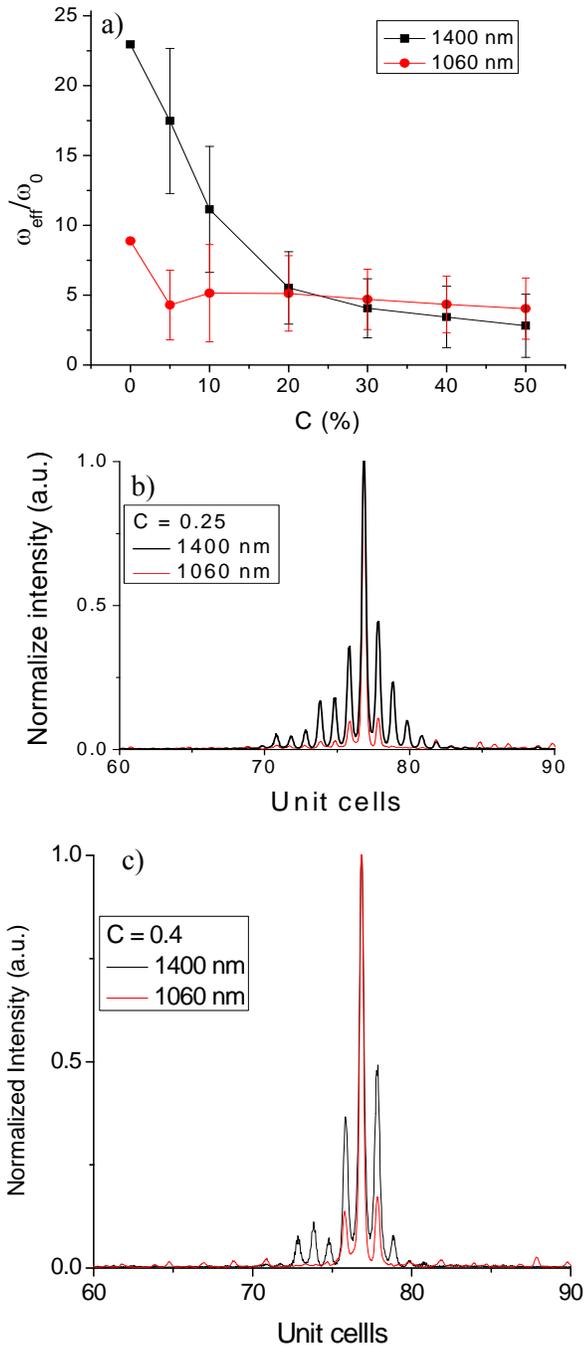

**Figure 5.** (Color online) a) Variation in the ensemble averaged (over 100 realizations) effective width (along with the ssd) of the output beam intensity versus levels of disorder for an input Gaussian beam (FWHM 8 μm) after propagation through 10 mm for two wavelengths: one at a wavelength (λ = 1400 nm) outside the bandgap window and the other, which is near the band-edge (λ = 1060 nm). Defect mode profiles by considering a defect waveguide at 77$^{th}$ unit cell of width 3.7 μm (different from the uniformly chosen width of 3 μm for all other waveguides) for these wavelengths for $C$ valus b) 0.25 and c) 0.40, respectively.

## 4. Conclusions

To conclude, we have studied the significance of a prominent underlying photonic bandgap spawned by the chosen waveguide lattice at the operating wavelength in the context of transverse localization of light. The results of our extensive numerical simulations reveal that on the background of a prominent bandgap, localization of light would occur even in the presence of relatively low level of disorder compared to its counterpart, in which PBG is absent. Our study also establishes the fact that disorder induced localization is achieved both inside as well as outside the underlying PBG. However, it is seen that existence of an underlying PBG for the waveguide lattice introduces a new tool to enhance the wavelength selectivity of the phenomenon in such 1D waveguide lattice; one could easily control the spectral window of light localization for specific applications including random lasing [20]. Hence, we envision that these results should be of interest in designing disordered lasers and other applications involving localization in imperfect lattice structures that spawn PBG.

## Acknowledgements

SG acknowledges the financial support by Department of Science and Technology, India as a INSPIRE Faculty Fellow [IFA-12; PH-13].
This work also relates to Department of the Navy Grant N62909-10-1-7141 issued by Office of Naval Research Global, which partially supported this work. The United States Government has royalty-free license throughout the world in all copyrightable material contained herein. This work was partially supported by UM.C/HIR/MOHE/SC/01.

**References**

[1] De Raedt H, Lagendijk A and De Vries P, 1989 *Phys. Rev. Lett.* **62**, 47
[2] John S, 1987 *Phys. Rev. Lett.* **58**, 2486
[3] Garanovich L, Longhi S, Sukhorukov A A and Kivshar Y S, 2012 *Phys. Rep.* **518**, 1
[4] Ghosh S, Psaila N D, Thomson R R, Pal B P, Varshney R K and Kar A K, 2012 *Appl. Phys. Lett.* **100**, 101102
[5] Martin L, Di Giuseppe G, Perez-Leija A, Keil R, Dreisow F, Heinrich M, Nolte S, Szameit A, Abouraddy A F, Christodoulides D N and Saleh B E A, 2011 *Opt. Exp.* **19**, 13636
[6] Schwartz T, Bartal G, Fishman S and Segev M, 2007 *Nature* **446**, 52
[7] Lahini Y, Avidan A, Pozzi F, Sorel M, Morandotti R, Christodoulides D N and Silberberg Y, 2008 *Phys. Rev. Lett.* **100**, 013906
[8] Pertsch T, Peschel U, Kobelke J, Schuster K, Bartelt H, Nolte S, Tünnermann A and Lederer F, 2004 *Phys. Rev. Lett.* **93**, 053901
[9] Ghosh S, Agrawal G P, Pal B P and Varshney R K, 2011 *Opt. Comm.* **284**, 201
[10] Toninelli C, Vekris E, Ozin G A, John S and Wiersma D S, 2008 *Phys. Rev. Lett.* **101**, 123901
[11] Vlasov Y A, Kaliteevski M A and Nikolaev V V, 1999 *Phys. Rev. B* **60**, 1555
[12] Kaliteevski M A, Beggs D M, Brand S, Abram R A and Nikolaev V V, 2006 *Phys. Rev. B* **73**, 033106
[13] García P D, Sapienza R, Froufe-Pérez L S and López C, 2009 *Phys. Rev. B* **79**, 241109(R)
[14] Leonetti M, Conti C and Lopez C, 2011 *Nature Photon.* **5**, 615
[15] Ghosh S, Pal B P and Varshney R K, 2012 *Opt. Comm.* **285**, 2785
[16] Ghosh S, Varsheny R K and Pal B P, 2013 *Laser Phys. Lett.* **10,** 085002
[17] Jović D M, Belić M R and Denz C, 2011 *Phys. Rev. A* **84**, 043811
[18] Szameit A, Kartashov Y V, Zeil P, Dreisow F, Heinrich M, Keil R, Nolte S, Tünnermann A, Vysloukh V A and Torner L, 2010 *Opt. Lett.* **35**, 8
[19] García P D, Smolka S, Stobbe S and Lodahl P, 2010 *Phys. Rev. B* **82**, 165103
[20] Redding B, Choma M A and Cao H, 2012 *Nature Photon*. **6**, 355